\documentclass[prb,a4paper,showpacs,twocolumn]{revtex4}
\usepackage{amsmath}
\usepackage{amssymb}
\usepackage{amsfonts}
\usepackage{graphicx}
\usepackage[usenames,dvipsnames]{color}
\usepackage{epstopdf}
\usepackage{bm}

\DeclareMathOperator{\Tr}{Tr} 

\DeclareMathOperator{\sgn}{sgn}

\DeclareMathOperator{\erfi}{erfi}
\DeclareMathOperator{\erf}{erf}

\newcommand{\bb}{\begin{equation}}
\newcommand{\ee}{\end{equation}}

\begin{document}

\title{Tunneling density of states in quantum dots with anisotropic exchange}

\author{A.U. Sharafutdinov and I.S. Burmistrov}
\affiliation{L.D. Landau Institute for Theoretical Physics RAS,
Kosygina street 2, 119334 Moscow, Russia}
\affiliation{Moscow Institute of Physics and Technology, 141700 Moscow, Russia}

\date{\today}

\begin{abstract}
We reexamine the tunneling density of states in quantum dots and nanoparticles within the model which is extension of the universal Hamiltonian to the case of uniaxial anisotropic exchange. We derive the exact analytical result for the tunneling density of states in the case of arbitrary single-particle energy spectrum.  We find that, similar to the case of the isotropic exchange, the tunneling density of states as a function of energy has the maximum due to a finite value of the total spin of the ground state near the Stoner instability. We demonstrate that there are no additional extrema which have been predicted on the basis of perturbative analysis [M.N. Kiselev and Y. Gefen, Phys. Rev. Lett. {\bf 96}, 066805 (2006)].
\end{abstract}


\maketitle


\section{Introduction}               

Quantum dots host rich physics that has been attracting experimental and theoretical interest for many years [\onlinecite{Alhassid2000,Wiel,ABG,Hanson,Ullmo2008}]. In the metallic regime when the Thouless energy ($E_{\rm Th}$) is much larger than the mean single-particle level spacing ($\delta$), $E_{\rm Th}/\delta \gg 1$, electrons in quantum dots can be universally described by an effective zero-dimensional Hamiltonian [\onlinecite{KAA}]. The advantage of this so-called universal Hamiltonian is the reduction of a set of matrix elements which describe electron-electron interaction in the single-particle basis to three parameters. They are the charging energy ($E_c$), the ferromagnetic Heisenberg exchange ($J>0$) and the Cooper-channel interaction. 
Typically, the charging energy is large, $E_c\gg \delta$, and suppresses a real electron tunneling through a quantum dot at low temperatures $T\ll E_c$: the phenomenon known as the Coulomb blockade [\onlinecite{CB}]. It can be seen as suppression of the tunneling density of states at low energies [\onlinecite{KamenevGefen1996,SeldmayrLY}]. 

Contrary to the charging energy, the exchange interaction typically is small, $J\leqslant \delta$. 
Provided the size ($L$) of a quantum dot is larger than the Fermi wave length, the exchange interaction can be estimated as $J/\delta = - F_0^\sigma$ where $F_0^\sigma$ denotes bulk value of the Fermi-liquid triplet channel interaction parameter. In bulk materials at $F_0^\sigma=-1$ there is the Stoner instability corresponding to the quantum phase transition between a paramagnet and a ferromagnet.  Hence the ground state in quantum dots is expected to become fully spin polarized at $J=\delta$. Surprisingly, an interesting regime with a finite total spin in the ground state is possible in quantum dots [\onlinecite{KAA}]. For the equidistant single-particle spectrum a partial spin polarization can be found at $\delta/2 \lesssim J<\delta$. With increase of $J/\delta$ towards the Stoner instability, the total spin in the ground state increases monotonously. This phenomenon of subsequent transitions between the ground states with different values of the total spin is termed as the mesoscopic Stoner instability [\onlinecite{KAA}]. At $J=\delta$ all electrons in a quantum dot become spin polarized. The mesoscopic Stoner instability is restricted to systems of a finite size and does not survive in the thermodynamic limit $\delta \to 0$.

The finite total spin of the ground state yields the Curie type behavior of the static spin susceptibility and its moments near the Stoner instability [\onlinecite{BGK1,BGK2,Saha2012,LSB,SLB}]. Interestingly, existence of the mesoscopic Stoner instability can be seen in the low temperature electron transport through a quantum dot. The finite total spin in the ground state leads to an additional nonmonotonicity (in comparison with nonmonotonicities due to Coulomb blockade) of the energy dependence of the tunneling density of states [\onlinecite{KiselevGefen,BGK1,BGK2}] and to enhancement of the shot noise [\onlinecite{Koenig2012}].

It is worthwhile to compare the case of isotropic (Heisenberg) exchange with the case of Ising exchange. For example, the latter can be realized in a two-dimensional quantum dot in the presence of strong spin-orbit coupling.
A spin-orbit coupling invalidates the universal Hamiltonian description of a quantum dot since
fluctuations of the interaction matrix elements cannot be neglected even in the metallic regime, $\delta/E_{\rm Th}\ll 1$ [\onlinecite{AlhassidSO,SOinQD}]. In case of a two-dimensional quantum dot only in-plane components of the spin are mixed with the orbital degrees of freedom whereas the perpendicular component of the total spin is conserved. If parameters of a quantum dot satisfy the following condition, $(\lambda_{SO}/L)^2 \gg (E_{\rm Th}/\delta)(L/\lambda_{SO})^4 \gg 1$, where $\lambda_{SO}$ stands for a spin-orbit length, the low energy description can be given by the universal Hamiltonian with the Ising exchange ($J_z>0$) [\onlinecite{AlhassidSO,AF2001}]. In this case there is no mesoscopic Stoner instability for the equidistant single-particle spectrum [\onlinecite{KAA}]. Since the total spin in the ground state is zero for all $J_z<\delta$, the tunneling density of states is almost independent of $J_z$ [\onlinecite{Boaz}].

The simplest way to incorporate Heisenberg and Ising interactions is to consider the universal Hamiltonian with an uniaxial anisotropy of exchange interaction. Albeit this model is not fully microscopically justified it can be relevant for nanometer-scale ferromagnetic nanoparticles. We note that significant anisotropy of an exchange interaction was revealed in experiments on tunneling spectra in such nanoparticles [\onlinecite{Gueron1999}]. The model which 
resembles the universal Hamiltonian with anisotropic exchange allows to explain the main features of experimentally measured excitation spectra [\onlinecite{Canali}]. Anisotropic exchange interaction in nanoparticles can be caused by   bulk magnetocrystalline, surface or shape anisotropy. The presence of spin-orbit scattering results in large mesoscopic fluctuations of the anisotropic part of the exchange interaction [\onlinecite{UB2005, BG2005}]. In quantum dots 
anisotropic exchange interaction can be induced by ferromagnetic leads [\onlinecite{Misiorny2013}]. 

The tunneling density of states for the universal Hamiltonian with uniaxial anisotropy of exchange interaction was studied in Ref. [\onlinecite{KiselevGefen}] by means of the perturbation expansion near the Ising case. It was found that anisotropic exchange interaction induces reentrant behavior (with two maxima and minimum) of the tunneling density of states as a function of energy. In contrast, energy dependence of the tunneling density of states  is monotonous in the case of Ising exchange [\onlinecite{Boaz}] and has the single maximum in the case of Heisenberg exchange [\onlinecite{BGK1,BGK2}]. Thus the result of Ref. [\onlinecite{KiselevGefen}] indicates interesting physics due to finite spin in the ground state in the presence of anisotropic exchange. However, this expectation does not supported by recent calculations of the spin susceptibilities for the model with anisotropic exchange [\onlinecite{SLB}]. The spin in the ground state monotonously reduces with change of the anisotropy from zero (in the case of Heisenberg exchange) to  the maximal value (in the case of Ising exchange).

In this paper we reexamine the problem of calculation of the tunneling density of states for the universal Hamiltonian extended to the case of exchange interaction with uniaxial anisotropy. Within this model we derive exact analytical result for the tunneling density of states. We analyze this exact result in the cases of zero temperature and
for temperatures larger than the mean single-particle level spacing. We demonstrate that 
\begin{itemize}
\item[(i)] similarly to the case of isotropic exchange, in vicinity of the Stoner instability the tunneling density of states  as a function of energy has the single maximum due to presence of the finite total spin in the ground state; 
\item[(ii)] there are no additional extrema in the energy dependence of the tunneling density of states contrary to predictions of perturbative analysis of Ref. [\onlinecite{KiselevGefen}].
\end{itemize} 

In our analysis, we use the following standard simplifications. We do not consider interaction in the Cooper channel which drives superconducting correlations in quantum dots [\onlinecite{SCinQD}]. This is allowed in the case of repulsive Cooper channel interaction [\onlinecite{KAA}]. Although our exact analytical result for the tunneling density of states in the case of uniaxial anisotropic exchange interaction is valid for an arbitrary single-particle spectrum, in its analysis we avoid consideration of randomness of single-particle levels. As we mentioned above in this case one needs to take into account corrections to the zero-dimensional Hamiltonian which stem from fluctuations of the matrix elements of the electron-electron interaction [\onlinecite{Altshuler1997,Mirlin1997}] in spite of the metallic regime, $\delta/E_{\rm Th}\ll 1$. In the case of isotropic exchange interaction these corrections are negligible but results in rich physics beyond the zero-dimensional approximation [\onlinecite{Ullmo2008}]. Also in the case of isotropic exchange interaction [\onlinecite{BGK2}] the fluctuations of single-particle levels do not affect strongly the tunneling density of states.

The outline of the paper is as follows. We start from definition of the model Hamiltonian and partial disentangling of spin and charge degrees of freedom (Sec. \ref{sec:formalism}). In Sec. \ref{Sec:TDOSexact} we derive the exact analytical expression for the tunneling density of states. Zero temperature behavior of the tunneling density of states is explored in Sec. \ref{Sec:ZeroT}. The temperatures well above the mean single-particle level spacing are considered in Sec. \ref{Sec:TDOS_T}. We conclude the paper with the summary (Sec. \ref{sec:dc}).

\section{Formalism \label{sec:formalism}}

\subsection{Hamiltonian}
\label{AUH}

We consider the following model:
\begin{equation}
H =H_0+H_{C}+H_{S} .
\label{ham}
\end{equation}
Here $H_0$ describes noninteracting electrons,
\begin{equation}
H_0=\sum_{\alpha}\sum_{\sigma=\pm}\epsilon_{\alpha} a^{\dag}_{\alpha\sigma}a_{\alpha\sigma} .
\end{equation}
It involves the spin-independent single-particle energy levels $\epsilon_{\alpha}$
and the single-particle creation ($a^{\dag}_{\alpha\sigma}$) and annihilation ($a_{\alpha\sigma}$) operators. The next term $H_C$ in the Hamiltonian \eqref{ham} takes into account the effect of direct Coulomb interaction among electrons in zero-dimensional approximation,
\begin{equation}
H_C=E_c(\hat{n}-N_0)^2 , \qquad \hat{n} =
\sum_{\alpha,\sigma}a^{\dag}_{\alpha,\sigma}a_{\alpha,\sigma} , 
\label{Hc}
\end{equation}
where $N_0$ denotes the background charge. The last term $H_S$ describes ferromagnetic anisotropic exchange interaction ($J_\perp>0$, $J_z>0$) 
\begin{equation}
\begin{split}
H_S &=-J_{\perp}(\hat{S}_x^2+\hat{S}_y^2)-J_z\hat{S}_z^2 , \\ 
\hat{\bm{S}}& =\frac{1}{2} \sum_{\alpha,\sigma\sigma^\prime} a^\dag_{\alpha\sigma}\bm{\sigma}_{\sigma\sigma^\prime}a_{\alpha\sigma^\prime}  .
\end{split} 
\label{Hs}
\end{equation}
Here $\bm{\sigma}=\{\sigma^x,\sigma^y,\sigma^z\}$ stands for the standard Pauli matrices. The exchange part $H_S$ of the Hamiltonian \eqref{ham} interpolates between the Heisenberg exchange, $J_\perp=J_z$ and the Ising exchange, $J_\perp=0$. In both cases, Eq. \eqref{ham} reduces to the universal Hamiltonian [\onlinecite{KAA}]. The Hamiltonian \eqref{ham} with the Ising exchange is used for description of lateral quantum dots with strong spin-orbit coupling [\onlinecite{AF2001,AlhassidSO}].

\subsection{Partial disentangling of spin and charge}

Our aim is to compute the tunneling density of states $\nu(\varepsilon)$ for the Hamiltonian \eqref{ham}. It 
can be conveniently expressed via single-particle Green's function in the Matsubara time domain
~[\onlinecite{MatveevAndreev}]
\begin{equation}
\nu(\varepsilon) = -\frac{1}{\pi} \cosh \frac{\beta \varepsilon}{2} \int\limits_{-\infty}^\infty dt\, e^{i\varepsilon t}  
\sum\limits_{\alpha,\sigma} G_{\alpha,\sigma\sigma }\left ( it+\beta/2\right ) , 
\label{EqTDOSdef}
\end{equation} 
where $\beta=1/T$. In the Lagrangian formalism, the Matsubara Green's function (matrix in the spin space) can be written as
\begin{gather}
G_{\alpha}(\tau_1,\tau_2) = -\frac{\mathcal{T}}{Z}  \int \mathcal{D}[\overline{\Psi},\Psi,\phi,\bm{\theta}] \Psi_{\alpha}(\tau_1)
\overline{\Psi}_{\alpha}(\tau_2)\, e^{-S_{\rm tot}} , \notag \\
Z = \int\mathcal{D}[\overline{\Psi},\Psi,\phi,\bm{\theta}]\, e^{-S_{\rm tot}} .
\label{eq:m:gf}
\end{gather}
Here $\mathcal{T}$ denotes the time ordering and $S_{\rm tot}$ is the imaginary time action for the Hamiltonian~\eqref{ham} after the Hubbard-Stratonovich transformation:
\begin{align}
S_{\rm tot} = &  \int\limits_0^\beta   d\tau  \Biggl \{ \sum_{\alpha}
\overline{\Psi}_{\alpha} \left [ \partial_\tau - \epsilon_\alpha +\mu  +i \phi+\frac{\bm{\sigma}\cdot \bm{\theta}}{2} \right ] \Psi_{\alpha}  \notag
\\
& + \frac{\theta_x^2+\theta_y^2}{4J_\perp}+\frac{\theta_z^2}{4J_z} +\frac{\phi^2}{4E_c}-i N_0 \phi  \Biggr \} .
\label{eq:Stot}
\end{align}
We have introduced the chemical potential $\mu$  and the Grassmann variables $\overline{\Psi}_{\alpha} = (\bar\psi_{\alpha\uparrow},\bar\psi_{\alpha\downarrow})^T, \Psi_{\alpha} = (\psi_{\alpha\uparrow},\psi_{\alpha\downarrow})$ to represent electrons on the dot. The scalar $\phi$ and vector $\bm{\theta}$ bosonic fields
were used to decouple the direct Coulomb and exchange interactions, respectively. We start from performing a gauge transformation in the charging sector by splitting the field $\phi(\tau)$ as
\begin{equation}
\phi(\tau) = \tilde{\phi}(\tau) + \frac{2\pi m}{\beta} + \phi_0, \, \int\limits_0^\beta d\tau\, \tilde{\phi}(\tau)=0 , \, |\phi_0|\leqslant \pi T 
\end{equation}
with integer $m$. The part $\tilde\phi(\tau)+2\pi mT$ of $\phi(\tau)$ can be gauged away (see Refs.~[\onlinecite{KamenevGefen1996,EfetovTscherisch,KiselevGefen,SeldmayrLY,Boaz}] for details). The Green's function \eqref{eq:m:gf} becomes 
\begin{gather}
{G}_{\alpha}(\tau_1,\tau_2) =  \int\limits_{-\pi T}^{\pi T}\frac{d\phi_0}{2\pi T} \,\frac{\mathcal{Z}(\phi_0)}{Z}
{D}(\tau_{12},\phi_0) \, \mathcal{G}_{\alpha}(\tau_{12},\phi_0),
\label{CSSep1} \\
Z = \int\limits_{-\pi T}^{\pi T}\frac{d\phi_0}{2\pi T} \, {D}(0,\phi_0) \mathcal{Z}(\phi_0) , \label{CSSep1Z}
\end{gather}
where $\tau_{12}\equiv\tau_1-\tau_2$. The so-called Coulomb-boson propagator reads
\begin{equation}
D(\tau,\phi_0) =  \sum_{k\in \mathbb{Z}}  e^{-E_c|\tau|(1-|\tau| T)+i\phi_0 (\beta k+\tau)-\beta E_c(k-N_0+\tau T)^2}.
\label{CB}
\end{equation}
The Green's function $\mathcal{G}_{\alpha}(\tau_{12},\phi_0)$ corresponds to the action $S_{\rm tot}$
with $\phi$ substituted by $\phi_0$ in the first line of Eq. \eqref{eq:Stot} and by $0$ in the second line. 
Thus, $\mathcal{G}_{\alpha}(\tau_{12},\phi_0)$ can be formally considered as the single-particle Green's function for the Hamiltonian $\mathcal{H} = \mathcal{H}_0 +H_S$ where $\mathcal{H}_0$ is given by $H_0$ (see Eq.~\eqref{ham}) in which $\epsilon_{\alpha,\sigma}$ is replaced by $\tilde{\epsilon}_{\alpha,\sigma}=\epsilon_{\alpha,\sigma} -\mu+i \phi_0$. We emphasize that the  charge and spin degrees of freedom are not fully disentangled. The remnant trace of $H_C$ is encoded in $\phi_0$ which leads to a small imaginary shift of the chemical potential. 
We remind that the grand partition function for the Hamiltonian \eqref{ham} is given as [\onlinecite{AlhassidRupp,SLB}]
\begin{align}
Z & =  \sum_{n_\uparrow,n_\downarrow}  Z_{n_\uparrow}Z_{n_\downarrow} e^{-\beta E_c(n-N_0)^2+\beta \mu n+ \beta J_\perp m(m+1)}\notag \\
& \times \sgn (2m+1) \sum_{l=-|m+1/2|+1/2}^{|m+1/2|-1/2} e^{\beta (J_z-J_\perp) l^2} ,
\end{align}
where $n_{\uparrow,\downarrow} = n/2\pm m$, and $Z_n$ is the Darwin-Fowler integral:
\begin{equation}
Z_n  = \int\limits_0^{2\pi} \frac{d\theta}{2\pi} e^{-i \theta n} \prod\limits_\gamma \left (1+e^{-\beta \epsilon_\gamma+i\theta} \right ).
\end{equation}

\section{Exact expression for the tunneling density of states\label{Sec:TDOSexact}} 

\subsection{Wei-Norman-Kolokolov transformation}

In the Hamiltonian formalism $\mathcal{G}_{\alpha}(\tau_{12})$ can be written as
\begin{equation}
\mathcal{G}_{\alpha}(\tau) =\frac{1}{\mathcal{Z}} \begin{cases}
- \mathcal{K}_\alpha(-i\tau,-i\tau+i\beta) , & \quad \tau > 0 ,\\
\mathcal{K}_\alpha(-i\tau-i\beta,-i\tau) , & \quad \tau \leqslant 0 ,
\end{cases}
\label{eq:GF:K:def}
\end{equation}
where $\mathcal{Z} = \exp(-\beta \mathcal{H})$ and 
\begin{equation}
 \mathcal{K}_{\alpha,\sigma_1,\sigma_2}(t_+,t_-) = \Tr \, e^{-i t_+ \mathcal{H}} a^\dag_{\alpha\sigma_1}
 e^{i t_- \mathcal{H}} a_{\alpha\sigma_2} .
 \label{eq:K:ham}
\end{equation}
Using the commutativity of $\mathcal{H}_0$ and $H_S$ we can split the evolution operator for $\mathcal{H}$ into two parts, $\exp(i t \mathcal{H}) = \exp(i t \mathcal{H}_0) \exp(i t{H}_S)$.  Next we apply the Hubbard-Stratonovich transformation to get rid of terms of the fourth order in electron operators in the exponent $H_S$: 
\begin{align}
e^{\mp it H_S}
 & = \lim_{N\rightarrow \infty}\int \Bigl[\prod_{n=1}^{N}d\bm\theta_n\Bigr] \prod_\alpha \mathcal{T} e^{it{\bm\theta_n \bm s_\alpha}/N}\notag \\
 & \times  \exp \left [ \pm \frac{i \Delta}{4}\sum\limits_{n=1}^{N}\left ( \frac{\theta^2_{x,n}+\theta^2_{y,n}}{J_\perp}+\frac{\theta^2_{z,n}}{J_z}\right )\right ] ,
\label{evol}
\end{align}
where $\Delta=t/N$. Here and further on we omit normalization factors. They will be restored in the final result. In what follows we shall concentrate on evaluation of $\mathcal{K}_{\alpha}(t_+,t_-)$. The corresponding partition function $\mathcal{Z}$ has been already computed in Ref. [\onlinecite{SLB}]. To make further progress with Eq. \eqref{evol} we apply the Wei-Norman-Kolokolov transformation [\onlinecite{WeiNorman,Kolokolov}] which allows us to rewrite $\mathcal{T}$-exponent as a product of usual exponents:
\begin{align}
& \mathcal{T}e^{i\Delta{\bm\theta_n \bm s_\alpha}}  = 
e^{p s_\alpha^{-p} \kappa_{p,N}^p} \exp \left (i s_\alpha^z \Delta \sum_{n=1}^N \rho_{p,n} \right ) \notag \\
& \hspace{1cm} \times
\exp \left ( i  s_\alpha^p \Delta \sum_{n=1}^N \kappa_{p,n}^{-p}  \prod_{j=1}^n e^{- i p \Delta \rho_{p,j}}\right ) ,
\label{eq:HBt}
\end{align}
where $s_\alpha^p = s_\alpha^x+ i p s_\alpha^y$. Equation \eqref{eq:HBt} is valid for both $p=\pm$. We use the initial condition $\kappa_{p,1}^p=0$. The variables $\bm{\theta}$ can be expressed via
new variables $\rho_p, \kappa_p^p$ and $\kappa_p^{-p}$ as follows:
\begin{align}
\frac{\theta_{x,n}-i p \theta_{y,n}}{2}
 & =\kappa^{-p}_{p,n} ,
\,\, \theta_{z,n}  =\rho_{p,n}-\kappa_{p,n}^{-p}(\kappa_{p,n}^p+\kappa_{p,n-1}^p) ,
\notag 
\\
\frac{\theta_{x,n}+i p \theta_{y,n}}{2}
& =\frac{\kappa^p_{p,n}-\kappa^p_{p,n-1}}{ip\Delta}+\frac{\rho_{p,n}(\kappa_{p,n}^p+\kappa_{p,n-1}^p)}{2}
\notag 
\\
&  -\frac{(\kappa_{p,n}^p+\kappa_{p,n-1}^p)^2}{4} \kappa_{p,n}^{-p}  
 .
\label{WNK}
\end{align}
A few comments are in order here. The vector $\bm{\theta}_n$ in Eq. \eqref{eq:HBt} is supposed to be real but the transformation \eqref{WNK} suggests that it is complex. This corresponds to rotation of the contour of integration in Eq. \eqref{eq:HBt}. To preserve the number of independent variables we chose $\rho_{p,n}$ to be purely imaginary, $\rho_{p,n}=-\rho_{p,n}^*$, and $\kappa_{p,n}^+$ and $\kappa_{p,n}^-$ to be complex conjugated, $\kappa_{p,n}^+ = (\kappa_{p,n}^-)^*$. The transformation \eqref{WNK} assumes that the quantity $(\kappa_{p,N}^p+\kappa_{p,N-1}^p)/2$ corresponds to $\kappa_p^p(t)$ in the continuous limit. In general, one can use any of discrete representations of $\kappa_p^p(t)$ of the form $\nu\kappa_{p,N}^p+(1-\nu)\kappa_{p,N-1}^p$ with $0\leqslant \nu \leqslant 1$. However, the symmetric one is special since for the choice $\nu=1/2$ it is sufficient to work with Eq. \eqref{evol} to the first order in $\Delta$. The Jacobian of the transformation \eqref{WNK} is given as  $\exp(i p \Delta\sum_{n=1}^{N} \rho_{p, n}/2)$ [\onlinecite{Kolokolov}].

Rewriting two exponents in Eq. \eqref{eq:K:ham} with the help of representation \eqref{eq:HBt}, we obtain
\begin{align}
\mathcal{K}_{\alpha \sigma_1\sigma_2} & = \prod_{p=\pm}\Biggl \{  \prod_{n_p=1}^{N_p} \int d\kappa_{p,n_p}^p d\kappa_{p,n_p}^{-p} d\rho_{p,n_p} \exp\Biggl [ \frac{i p \Delta \rho_{p,n_p}}{2}  \notag \\
& \times \Biggl (1  -\frac{\rho_{p,n_p}}{2J_z}  - \frac{\varkappa }{J_\perp}\kappa_{p,n_p}^{-p} \bigl (\kappa_{p,n_p}^p+\kappa_{p,n_p-1}^p\bigr ) \Biggr )\notag \\
& + \frac{i p \Delta}{4J_\perp} \bigl (\kappa_{p,n_p}^{-p}\bigr )^2\bigl (\kappa_{p,n_p}^p+\kappa_{p,n_p-1}^p\bigr )^2\notag \\
& -\frac{\kappa_{p,n_p}^{-p}}{J_\perp}(\kappa_{p,n_p}^p-\kappa_{p,n_p-1}^p) \Biggr ]
\Biggr \} \prod\limits_{\gamma\neq\alpha}
\Tr [\mathcal{A}^{(+)}_\gamma \mathcal{A}^{(-)}_\gamma] \notag \\
& \times  \Tr [\mathcal{A}^{(+)}_\alpha a^\dag_{\alpha\sigma_1}\mathcal{A}^{(-)}_\alpha a_{\alpha\sigma_2}] .
\label{eq2}
\end{align}
Here the limit $N_p\to \infty$ is assumed. The quantity $\varkappa=1-{J_\perp}/{J_z}$ characterizes a deviation from the case of isotropic exchange.  

The single-particle operators  $\mathcal{A}^{(p)}_\alpha$ represent the evolution operators. In accordance with Eqs.  \eqref{eq:HBt} - \eqref{WNK} they are defined as follows
\begin{align}
\mathcal{A}^{(p)}_\alpha& =e^{-ipt_p \epsilon_\alpha n_\alpha} e^{p s^{-p}_\alpha \kappa_{p,N_p}^p}
\exp \left (i s_\alpha^z \Delta \sum_{n=1}^{N_p} \rho_{p,n}\right ) \notag \\
& \times
\exp \left [ i  s_\alpha^p \Delta \sum_{n=1}^{N_p} \kappa_{p,n}^{-p}  \exp\bigl (- i p \Delta \sum_{j=1}^n \rho_{p,j}\bigr )\right ] .
\end{align}
Due to non-zero value of parameter $\varkappa$ the action for $\kappa$'s in Eq. \eqref{eq2} is not Gaussian: there are forth order terms. To get rid of such terms, we introduce a set of auxiliary variables $\eta_{p,n_p}$ by employing the Hubbard-Stratonovich transformation:
\begin{gather}
\exp \left [ \frac{ip \Delta\varkappa}{4J_\perp} \bigl( \kappa^{-p}_{p,n_p}\bigr )^2 \bigl (\kappa_{p,n_p}^p+\kappa_{p,n_p-1}^p\bigr )^2 \right ] = \sqrt{\frac{ip\Delta \varkappa}{4\pi J_\perp}}\notag \\
 \times \int  d\eta_{p,n_p}  \exp \left ( \frac{ip \Delta\varkappa}{4 J_\perp} \eta_{p,n_p}^2 \right ) \notag \\
\times  \exp \left [  -\frac{ip \Delta\varkappa}{2 J_\perp} \eta_{p,n_p}\kappa_{p,n_p}^{-p}\bigl (\kappa_{p,n_p}^p+\kappa_{p,n_p-1}^p\bigr ) \right ] .
\end{gather}
Next following Ref. [\onlinecite{Kolokolov}], we introduce new variables
\begin{equation} 
\kappa_{p,n_p}^{-p}=\chi_{p,n_p}^{-p} e^{\alpha_{p,n_p}},\qquad
\kappa_{p,n_p}^{p}=\chi_{p,n_p}^{p} e^{\beta_{p,n_p}} ,
\label{eq:tr2}
\end{equation}
where 
\begin{equation}
\begin{split}
\beta_{p,n_p} & = - i p \Delta \varkappa \sum_{n=1}^{n_p} (\rho_{p,n}-\eta_{p,n}) , 
\\
\alpha_{p,n_p}  & =-\beta_{p,n_p} -\frac{i p\Delta \varkappa}{2}  (\rho_{p,n_p} -\eta_{p,n_p})  .
\end{split}
\label{Jacob}
\end{equation}
Such choice of $\alpha_{p,n_p}$ and $\beta_{p,n_p}$ allows us to remove terms of the second order in $\chi$'s and first order in $\rho$ in the action in Eq. \eqref{eq2}. It can be done within accuracy of the first order in $\Delta$.  We note that one needs to take into account the Jacobian of  the transformation \eqref{Jacob}, 
\begin{equation}
\mathcal{J}_p=\exp\Bigl [ - i p \Delta \varkappa \bigl (\rho_{p,n_p} -\eta_{p,n_p}\bigr )/2\Bigr ].
\end{equation} 

In the absence of magnetic field, $\mathcal{K}_{\alpha\uparrow\downarrow} = \mathcal{K}_{\alpha\downarrow\uparrow} = 0$ and $\mathcal{K}_{\alpha\uparrow\uparrow} = \mathcal{K}_{\alpha\downarrow\downarrow}$. Therefore, we concentrate on calculation of $K_{\alpha\uparrow\uparrow}$ below. After evaluation of the single-particle traces in the expression \eqref{eq2} we find
\begin{widetext}
\begin{gather}
\mathcal{K}_{\alpha\uparrow\uparrow}(t_+,t_-) =  \prod_{p=\pm}\Biggl \{ \prod_{n_p=1}^{N_p} \int d\chi_{p,n_p}^{p} d\chi_{p,n_p}^{-p} d\rho_{p,n_p}d\eta_{p,n_p} \, \exp \Biggl [\frac{i p \Delta}{2} \bigl [(1-\varkappa)\rho_{p,n_p}+\varkappa \eta_{p,n_p}\bigr ]-\frac{i p \Delta}{4J_z} \Bigl [\rho_{p,n_p}^2+\frac{\varkappa \eta_{p,n_p}^2}{1-\varkappa} \Bigr ]\notag \\
  -\frac{\chi_{p,n_p}^{-p}}{J_\perp}(\chi_{p,n_p}^p-\chi_{p,n_p-1}^p)\Biggr ]
\Biggr \}   e^{-2i\epsilon_\alpha t_+}\sum_{p=\pm}e^{i\epsilon_\alpha t_p}\exp \Bigl [ \frac{ip\Delta}{2}\sum\limits_{n_p=1}^{N_p}\rho_{p,n_p}\Bigr ]\prod\limits_{\gamma\neq\alpha}
\Biggl \{1+e^{-2i\epsilon_\gamma(t_+-t_-)}
+2e^{-i\epsilon_\gamma(t_+-t_-)}\notag \\
\times \cos\left (\frac{\Delta}{2}\sum_{p=\pm}\sum_{n_p=1}^{N_p}\rho_{p,n_p}\right )+ 
\prod_{p=\pm} e^{-ip \epsilon_\gamma t_p}
\exp\Bigl [ \frac{ip\Delta}{2}\sum\limits_{n_p=1}^{N_p}\rho_{p,n_p}\Bigr ] \Biggl ( p \chi_{p,N_p}^p \exp\Bigl [- i p \Delta \varkappa \sum\limits_{n_p=1}^{N_p} (\rho_{p,n_p}-\eta_{p,n_p})\Bigr ]\notag \\
+ i \Delta  \sum_{n_{-p}=1}^{N_{-p}} \chi_{-p,n_{-p}}^p \exp \Bigl [- i p \Delta \varkappa \sum\limits_{n=1}^{n_{-p}} (\rho_{-p,n}-\eta_{-p,n})+i p \Delta \sum\limits_{n=1}^{n_{-p}} \rho_{-p,n}\Bigr ] \Biggr )\Biggr \} .
\end{gather}
After integration over variables $\chi_{p,n_p}$ (see details in Appendix B of Ref. [\onlinecite{BGK2}]) we obtain
\begin{gather}
\mathcal{K}_{\alpha\uparrow\uparrow}(t_+,t_-) =  \prod_{p=\pm} \Biggl \{\prod_{n_p=1}^{N_p} \int d\rho_{p,n_p} d\eta_{p,n_p} 
 \, e^{\frac{i p \Delta}{2} [(1-\varkappa)\rho_{p,n_p}+\varkappa \eta_{p,n_p}]} e^{-\frac{i p \Delta}{4J_z} [\rho_{p,n_p}^2+\frac{\varkappa}{1-\varkappa} \eta_{p,n_p}^2] }\Biggr \} 
 \prod_\gamma\left (\oint\limits_{|z_\gamma|=1}\frac{i dz_\gamma}{2\pi z_\gamma^2}\right ) e^{-w_\alpha-2i\epsilon_\alpha t_+}
\notag \\
\times \sum_{p=\pm}e^{i\epsilon_\alpha t_p}e^{\frac{ip\Delta}{2}\sum_{n_p=1}^{N_p} \rho_{p,n_p}}
\exp \Biggl (- 2 v_\alpha \cos\left [\frac{\Delta}{2}\sum_{p=\pm} \sum_{n_p=1}^{N_p} \rho_{p,n_p}\right ] \Biggr )
\int\limits_0^\infty dy \, e^{-y}
 \exp \Biggl \{ -i J_\perp v_\alpha y \left ( \prod \limits_{p=\pm} e^{i \frac{p \Delta}{2} \sum_{n_p=1}^{N_p} \rho_{p,n_p}}\right ) \notag \\
\times
 \left ( \sum_{p=\pm} p\, e^{- i p \Delta \varkappa \sum_{n_p=1}^{N_p} (\rho_{p,n_p}-\eta_{p,n_p})}
 \Delta \sum_{n_p=1}^{N_p} e^{- i p \Delta \sum_{n=1}^{n_p} [(1-\varkappa)\rho_{p,n}+\varkappa\eta_{p,n}]}
 \right )
 \Biggr \} ,
 \label{eq:Zj1}
\end{gather}
\end{widetext}
where
\begin{equation}
\begin{split}
v_\alpha &= \sum_{\gamma\neq\alpha} z_\gamma  e^{-i\epsilon_\gamma(t_+-t_-)} ,  \\
w_\alpha & =\sum_{\gamma\neq\alpha} z_\gamma \left ( 1+e^{-2i\epsilon_\gamma(t_+-t_-)}\right ) .
\end{split}
\label{app:a:bd}
\end{equation}
Now it is convenient to switch to continuous representation. In order to transform  expression \eqref{eq:Zj1} into a more standard form, we introduce new variables:
\begin{equation}
\xi_{p}(t)=ip \int_0^t dt^\prime [(1-\varkappa) \rho_{p}(t^\prime)+\varkappa \eta_{p}(t^\prime)]+ \xi_{p}(0) ,
\end{equation}
satisfying the following relations:
\begin{gather}
\begin{split}
\sum_{p=\pm} p \Bigl [ \xi_p(0)-\varkappa \xi_p(t_p) & + i p \varkappa \int\limits_0^{t_p}dt \eta_p(t) \Bigr ] = 0 , \\
\sum_{p=\pm} \xi_p(t_p) & + 2 \ln (4v_\alpha y) = 0 .
\end{split}
\label{Eq:BC}
\end{gather}
Then after integration over variables $\eta_p$ and introduction of auxiliary variable $x$ we can write the functional integral for $K_{\alpha\uparrow\uparrow}$ as the integral of the Feynman-Kac type:
\begin{gather}
\mathcal{K}_{\alpha\uparrow\uparrow}= e^{-2i\epsilon_\alpha t_+}  \int\limits_{-\infty}^\infty dx\,  e^{-i J_z \varkappa  x^2 (t_+-t_-)}  \prod_{p=\pm}  \Biggl \{  \int \mathcal{D}[\xi_p]  \notag \\ 
\times
 e^{i p \int_0^{t_p} dt \mathcal{L}_p -(1-2i p x) \xi_p(0)/2}
\Biggr \}
  \prod_{\gamma\neq \alpha}\left (\oint\limits_{|z_\gamma|=1}\frac{i dz_\gamma}{2\pi z_\gamma^2}\right )
  \notag \\
  \times 
\int\limits_0^\infty \frac{dy}{4 y v_\alpha} \, e^{-y-w_\alpha - 2 v_\alpha \cosh [(\xi_{+}(t_+)-\xi_{-}(t_-))/2] }
\notag \\
\times \delta \Bigl (  \xi_+(t_+) +\xi_-(t_-) + 2 \ln (4v y)\Bigr )\notag \\
\times \sum\limits_{p=\pm}\Bigl [e^{(i\epsilon_\alpha-\varkappa x J_z+ip\varkappa J_z/4) t_p+[\xi_p(t_p)-\xi_p(0)]/2} \Bigr] .
 \label{eq:Zj4}
\end{gather}
Here the Lagrangians $\mathcal{L}_p$ are given as 
\begin{equation}
\mathcal{L}_p = \frac{1}{4J_\perp} \dot\xi_p^2 - \frac{J_\perp}{4} e^{-\xi_p} .
\label{eq:Lag}
\end{equation}
Now it is more convenient to rewrite Eq. \eqref{eq:Zj4} in the Hamiltonian representation: 
\begin{gather}
\mathcal{K}_{\alpha\uparrow\uparrow} = e^{-2i\epsilon_\alpha t_+}\prod_{\gamma\neq\alpha}\left (\oint\limits_{|z_\gamma|=1}\frac{i dz_\gamma}{2\pi z_\gamma^2}\right ) \int\limits_0^\infty \frac{dy}{4 y v_\alpha} \, e^{-y-w_\alpha}
\notag \\
\times
\int\limits_{-\infty}^\infty dx   \prod_{p=\pm}  \left \{ \int d\xi_p d\xi_p^\prime \, 
\, e^{-i J_z \varkappa  x^2  p t_p -(1-2i p  x) \xi_p^\prime/2}
\right \}
\notag \\
\times
\delta \left ( \sum_{p=\pm} \xi_p + 2 \ln (4v_\alpha y)\right )
e^{- 2 v_\alpha \cosh [(\xi_+-\xi_-)/2]}
\notag \\
\times
\langle \xi_+ | e^{-i \mathcal{H}_J t_+} | \xi_+^\prime \rangle 
\langle \xi_-^\prime | e^{i \mathcal{H}_J t_-} | \xi_- \rangle
\notag \\\times  
\sum\limits_{p=\pm}\Bigl [e^{i\epsilon_\alpha t_p}e^{\frac{\xi_p-\xi_p^\prime}{2}}e^{\frac{ip\varkappa J_z t_p}{4}}e^{-\varkappa x J_z t_p} \Bigr] .
 \label{eq:Zj5}
\end{gather}
The Hamiltonian of one-dimensional quantum mechanics corresponding to the Lagrangian \eqref{eq:Lag} reads [\onlinecite{Kolokolov}]
\begin{equation}
\mathcal{H}_J=-J_\perp\frac{\partial^2}{\partial\xi^2}+\frac{J_\perp}{4}e^{-\xi} .
\end{equation}
Its eigenvalues are given by $J \nu^2$ and eigenfunctions are spanned by the modified Bessel functions $K_{2i\nu}$ where $\nu$ is a real number:
\begin{equation}
\langle\xi |\nu\rangle = \frac{2}{\pi} \sqrt{\nu \sinh(2\pi \nu)} K_{2i\nu}(e^{-\xi/2}) .
\end{equation}

Using the following result (see formula 6.794.11 on p. 743 of Ref. [\onlinecite{GR}])
\begin{gather}
\int\limits_0^\infty d\nu \, \nu \sinh(2\pi \nu) K_{2i\nu}(2v_\alpha)K_{2i\nu}(e^{-\xi_+/2}) K_{2i\nu}(e^{-\xi_-/2})
\notag \\
= \frac{\pi^2}{16}\exp \left ( -\frac{1}{4v_\alpha} e^{-\frac{\xi_++\xi_-}{2}} -2 v_\alpha \cosh\frac{\xi_+-\xi_-}{2}
\right ) ,
\end{gather}
we can integrate over $y$, $\xi_+$, and $\xi_-$. Then we obtain [$\zeta =(\xi_-^\prime-\xi_+^\prime)/2$]
\begin{gather}
\mathcal{K}_{\alpha\uparrow\uparrow}= e^{-2i\epsilon_\alpha t_+} \prod_{\gamma\neq \alpha}\left (\oint\limits_{|z_\gamma|=1}\frac{i dz_\gamma}{2\pi z_\gamma^2}\right ) \frac{e^{-w_\alpha}}{v_\alpha}
\int\limits_{-\infty}^\infty dx d\zeta \, e^{\zeta/2} 
\notag \\ 
e^{-i J_z \varkappa x^2(t_+-t_-)}
\int\limits_0^\infty d\nu K_{2i\nu}(2v_\alpha) \int d\nu_1  \langle\nu |e^{\xi/2}|\nu_1\rangle
\sum\limits_{p=\pm} 
\notag \\
\times \Biggl \{ e^{(i\epsilon_\alpha-\varkappa x J_z+ip\varkappa J_z/4) t_p} 
e^{-i p J_\perp \nu_1^2 t_p +i p J_\perp \nu^2 t_{-p}}
\notag \\
\times
e^{2i x p \zeta- p \zeta/2} Q_{\nu\nu_1}(e^{p\zeta/2})
 \Biggr \} ,
\end{gather}
where
\begin{gather}
Q_{\nu_+\nu_-}(z) = z \int\limits_{-\infty}^\infty d\xi \, e^{-3\xi/2} \prod_{p=\pm} \langle \nu_p | \xi+2 p \ln z\rangle  .
\end{gather}

Now we use the following identity (see formula 6.576.4 on p. 676 of Ref. [\onlinecite{GR}])
\begin{gather}
\int\limits_0^\infty dx\, x^{-\lambda}K_\mu(ax)K_\nu(bx)= \frac{a^{-\nu+\lambda-1}b^\nu}{2^{2+\lambda}\Gamma(1-\lambda)}\prod\limits_{p,q=\pm} \Gamma\left (r_{pq}\right )
\notag\\
\times\,
{}_2F_{1}(r_{++},r_{-+},1-\lambda;1-{b^2}/{a^2}) ,
\end{gather}
where $r_{pq}=(1-\lambda+p \mu+q \nu)/2$, $\Gamma(x)$ stands for the Gamma function, and ${}_2F_{1}(a,b,c;z)$ denotes the hypergeometric function.
Then we obtain ($t_+-t_- = -i\beta$)
\begin{gather}
\mathcal{K}_{\alpha\uparrow\uparrow}(t_+,t_-)=\frac{e^{-2i\epsilon_\alpha t_+}
}{2\sqrt{\pi^3 \beta J_\perp}} \int\limits_{-\infty}^\infty dx d\zeta \, e^{2ix\zeta -\beta J_z \varkappa x^2}
\notag \\ \times
\int\limits_{-\infty}^\infty dh \, \sinh(h)  \prod_{\gamma \neq \alpha}\prod_{\sigma = \pm} (1+e^{-\beta \epsilon_\gamma - \sigma h}) \sum\limits_{p=\pm} \Bigl [e^{-(1+p)\zeta/2} \notag \\
\times 
 e^{i(\epsilon_\alpha-\varkappa x J_z +ipJ_z/4) t_p} 
\mathcal{W}(2h + ipJ_\perp t_p,\zeta, \beta J_\perp)\Bigr ] ,
\label{eq:K:fin}
\end{gather}
where the function $\mathcal{W}$ is defined as
\begin{align}
\mathcal{W}(x,y,z) & =\frac{1}{4\sinh y} \Biggl[\sum_{\sigma=\pm}\frac{\sigma \sqrt{\pi z}}{\sinh y}\erf\Bigl(\frac{x-2\sigma y}{2\sqrt{z}}\Bigr)\notag \\
& +4e^{-y}\exp\Bigl(-\frac{(x-2y)^2}{4z}\Bigr)\Biggr ] .
\end{align}
Here $\erf(z) = (2/\sqrt{\pi})\int_0^z dt\, \exp(-t^2)$ stands for the error function. Also we restored all necessary normalization factors.
Next integrating over $x$, $\zeta$, $h$ and $\phi_0$ in Eqs. \eqref{CSSep1} and \eqref{eq:GF:K:def}, we find the following result for the single-particle Green's function:
\begin{gather}
G_{\alpha\uparrow\uparrow}(\tau)=- \sum_{n_{\uparrow,\downarrow}\in \mathbb{Z}}e^{-\beta E_c(n-N_0)^2+\beta \mu n +\beta J_\perp m(m+1)}\notag \\ \times 
\frac{\sqrt{\beta\pi}}{8Z\sqrt{J_z-J_\perp}}e^{(J_z-J_\perp)\tau(1-\tau T)} \int \limits_{-\infty}^\infty db\, e^{\frac{-\beta b^2}{4(J_z-J_\perp)}} \notag \\ \times e^{-[\epsilon_\alpha-\mu + E_c(2n-2N_0+1)+J_\perp(m+1/4)+b/2]\tau}\Biggl\{e^{\beta b/2}\notag \\ \times \Upsilon(\beta b,2m+1)\Bigl [Z_{n_\uparrow}(\epsilon_\alpha)Z_{n_\downarrow}-Z_{n_\uparrow+1}Z_{n_\downarrow-1}(\epsilon_\alpha)\Bigr ] \notag \\ -\Upsilon(-\beta b,-2m) \Bigl [Z_{n_\uparrow}Z_{n_\downarrow}(\epsilon_\alpha)-Z_{n_\uparrow}(\epsilon_\alpha)Z_{n_\downarrow}\Bigr ] \Biggr\} .
\label{eq:F:G}
\end{gather}
Here $n_{\uparrow,\downarrow} = n/2\pm m$, and $Z_n(\epsilon_\alpha)$ is the integral of the Darwin-Fowler type:
\begin{align}
 Z_n(\epsilon_\alpha) = \int\limits_0^{2\pi} \frac{d\theta}{2\pi} e^{-i \theta n} \prod\limits_{\gamma \neq \alpha}\left (1+e^{-\beta \epsilon_\gamma+i\theta} \right ) ,
\end{align}
and the function
\begin{equation}
\Upsilon(z,x) = \frac{e^{(x-1)z/2}}{\sinh(z/2)} - \frac{\sinh(xz/2)}{x\sinh^2(z/2)} .
\end{equation}
Finally, substituting  the expression \eqref{eq:F:G} for the Green's function into Eq. \eqref{EqTDOSdef} and performing integration, we obtain the following result for the tunneling density of states for the Hamiltonian \eqref{ham}:
\begin{widetext}
\begin{align}
\nu(\varepsilon) &= \frac{1+e^{-\beta\varepsilon}}{Z} \sum_{n_\uparrow,n_\downarrow}  Z_{n_\uparrow}Z_{n_\downarrow} e^{-\beta E_c(n-N_0)^2+\beta \mu n+ \beta J_\perp m(m+1)}\sgn (2m+1) \sum_{l=-|m+1/2|+1/2}^{|m+1/2|-1/2} e^{\beta (J_z-J_\perp) l^2}
\sum\limits_\alpha  \notag \\
& \times
\Biggl \{  \delta \Bigl (\varepsilon - \epsilon_{\alpha}+\mu-E_c(2n-2N_0+1)-J_\perp(m+1/4)+(J_z-J_\perp)(l+1/4)\Bigr )  \frac{m-l}{m}\Biggl [\frac{Z_{n_\downarrow}(\epsilon_\alpha)}{Z_{n_\downarrow}} -  \frac{Z_{n_\uparrow}(\epsilon_\alpha)}{(2m+1)Z_{n_\uparrow}}\Biggr ] 
\notag \\
& +\delta \Bigl (\varepsilon - \epsilon_{\alpha}+\mu-E_c(2n-2N_0+1)+J_\perp(m+3/4)+(J_z-J_\perp)(l+1/4)\Bigr ) \frac{2m+2+2l}{2m+1}\frac{Z_{n_\uparrow}(\epsilon_\alpha)}{Z_{n_\uparrow}}
 \Biggr \} .
\label{TDoS}
\end{align}
\end{widetext}
This equation constitutes the main result of the present paper. We emphasize that our result
 is valid for an arbitrary single-particle spectrum $\{\epsilon_\alpha\}$. Each term in Eq. \eqref{TDoS}
corresponds to the tunneling of an electron with energy $\varepsilon$ and a given spin into (from) a single-particle level with energy $\epsilon_\alpha$. Each delta-function describes the energy conservation. The factor $Z_{n}(\epsilon_\alpha)/Z_{n}$ measures the probability that the single particle level with energy $\epsilon_\alpha$ is empty provided the total number of electrons is $n$. In the isotropic limit, $J_z=J_\perp$, Eq. \eqref{TDoS} coincides with the result obtained in Refs. [\onlinecite{BGK1,BGK2}]. In the case of Ising exchange, $J_\perp=0$, Eq. \eqref{TDoS} transforms into the result of Ref. [\onlinecite{Boaz}]. In the absence of exchange interaction, $J_z=J_\perp=0$, the result \eqref{TDoS} coincides with the expression found in Ref.
[\onlinecite{SeldmayrLY}]. 

Due to breaking degeneracy of many-particle spectrum by exchange anisotropy, each delta-peak for isotropic case is replaced by $2m+1$ peaks. Envelope of this set of peaks has the width of the order of $2m(J_z-J_\perp)$. As we will demonstrate below, this leads to smearing of the peak in the tunneling density of states in comparison with the case of isotropic exchange. 

By using the identities, $\sum_\alpha [Z_n-Z_n(\epsilon_\alpha)] = n Z_n$ and $Z_n=Z_n(\epsilon_\alpha)+e^{-\beta \epsilon_\alpha} Z_{n-1}(\epsilon_\alpha)$, one can check that Eq. \eqref{TDoS} fulfills the sum rule:
\begin{equation}
\int\limits_{-\infty}^\infty d\varepsilon \frac{\nu(\varepsilon)}{1+e^{\beta\varepsilon}} = T \frac{\partial \ln Z}{\partial \mu} .
\label{eq:sumrule}
\end{equation}

\section{Zero temperature analysis\label{Sec:ZeroT}}

\begin{figure}[t]
	\centerline{\includegraphics[width=8cm]{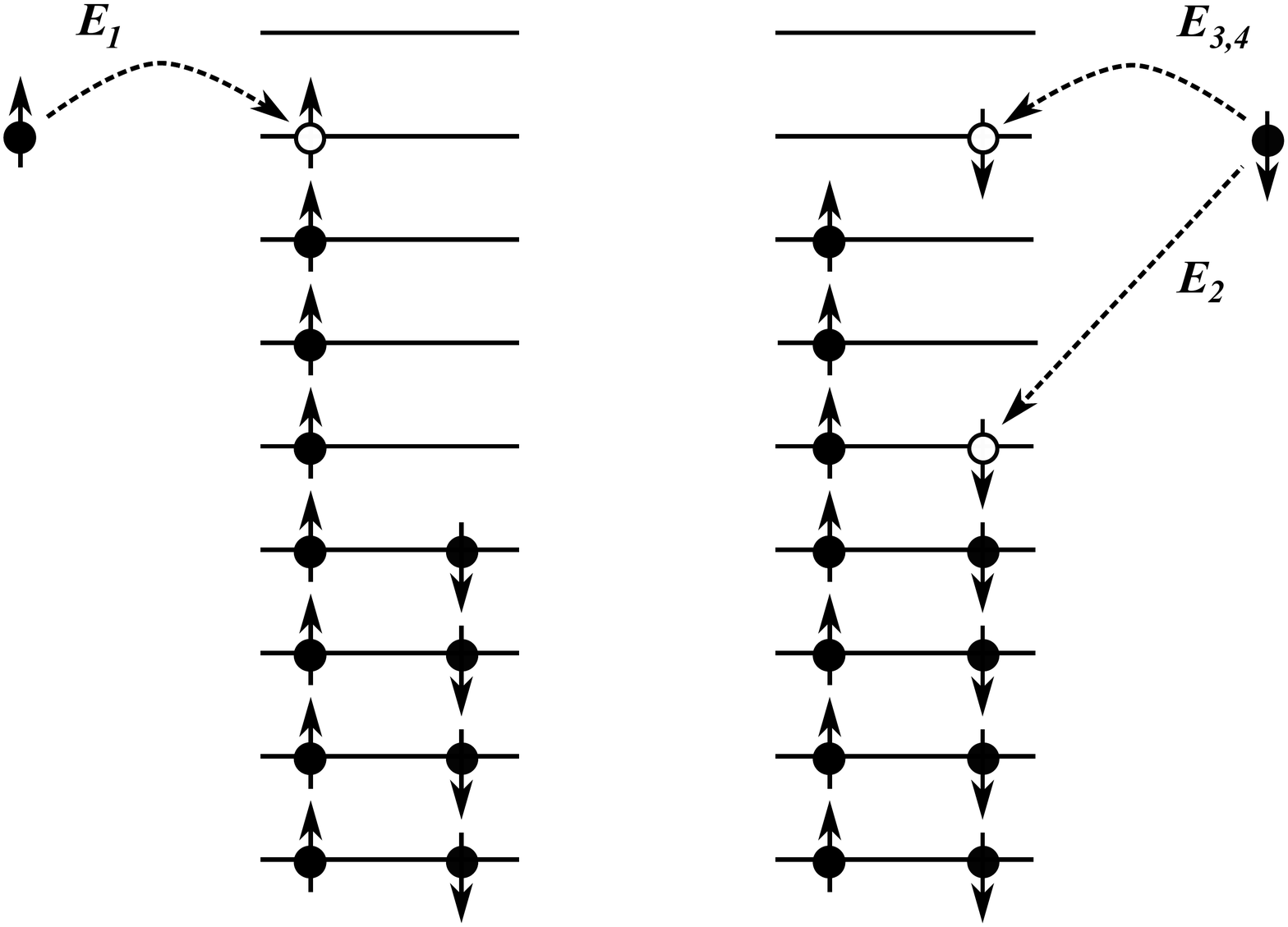}}
	\caption{(Color online) Tunneling of an electron with spin up (left) and spin down (right) into a quantum dot with a finite value of spin in the ground state.}
	\label{Fig:XXX}
\end{figure}

We start analysis of Eq. \eqref{TDoS} from the case of low temperatures $T\ll \delta$. For simplicity, we consider the case of Coulomb valley ($N_0$ is close to an integer). Then, at $T\ll \delta$ we can use the following relations:
\begin{equation}
	Z_n \approx e^{-\beta E_n^{(0)}}, \qquad \frac{Z_n(\epsilon_\alpha)}{Z_n} \approx  
	\Theta\bigl ( \epsilon_\alpha - E_n^{(0)} + E_{n-1}^{(0)}\bigr ) ,
\end{equation}
where $E_n^{(0)}$ stands for the ground state energy of $n$ spinless electrons and $\Theta(x)$ denotes the Heaviside step function. In the case of the equidistant spectrum it is equal to $E_n^{(0)}= \delta n (n-1)/2$. Let us assume that the ground state of the Hamiltonian \eqref{ham} with $J_z\geqslant J_\perp$ corresponds to the total spin $S$. Then in Eq. \eqref{TDoS}  one needs to take into account the contribution with $m=S$ and $l=\pm S$ 
only.  ïther contributions, e.g. with $m=S-1$ and $l=\pm(S-1)$, will be exponentially small. Hence, we find
\begin{align}
	\nu(\varepsilon) & =  
\sum\limits_{\epsilon_\alpha > \epsilon_{\frac{N_0}{2}-S}}  \delta \Bigl (\tilde\varepsilon_{\alpha}-J_z S -\frac{J_\perp}{2}+\frac{J_z}{4}\Bigr )	
\notag \\
	& 
	 - \frac{1}{2S+1} \sum\limits_{\epsilon_\alpha > \epsilon_{\frac{N_0}{2}+S}} \delta \Bigl (\tilde\varepsilon_{\alpha}- J_z S -\frac{J_\perp}{2}+\frac{J_z}{4}\Bigr )\notag \\
	&  + \frac{1}{2S+1}
	\sum\limits_{\epsilon_\alpha >\epsilon_{\frac{N_0}{2}+S}}\delta \Bigl (\tilde\varepsilon_{\alpha}+(2J_\perp-J_z)S +\frac{J_\perp}{2} +\frac{J_z}{4}\Bigr ) 
	\notag \\
	& 
	+  \sum\limits_{\epsilon_\alpha > \epsilon_{\frac{N_0}{2}+S}}
	\delta \Bigl (\tilde\varepsilon_{\alpha}+J_z S +\frac{J_\perp}{2} + \frac{J_z}{4}\Bigr )
	 .
	\label{TDoS-3}
\end{align}
Here $\tilde\varepsilon_\alpha = \varepsilon+\mu-E_c- \epsilon_{\alpha}$. It is convenient to rewrite $\epsilon_{\frac{N_0}{2}\pm S+1}$ as $\epsilon_{\frac{N_0}{2}\pm S+1} = E_{S\pm 1/2}-E_S$, where $E_S$ stands for the single-particle contribution to the energy of the ground spin with the spin $S$. Then let us introduce the following energies:
\begin{equation}
\begin{split}
	\mathcal{E}_1 & = E_{S+ 1/2}-E_S-J_z S - \frac{J_\perp}{2} -\frac{J_z}{4} , \\
	\mathcal{E}_2 & = E_{S-1/2}-E_S+J_z S  +\frac{J_\perp}{2}  - \frac{J_z}{4} , \\
	\mathcal{E}_3 & = E_{S+ 1/2}-E_S+J_z S  +\frac{J_\perp}{2} - \frac{J_z}{4} ,  \\ 
	\mathcal{E}_4 & = E_{S+ 1/2}-E_S+(J_z-2J_\perp) S - \frac{J_\perp}{2} -\frac{J_z}{4} .
\end{split}
\end{equation}
We note that the following inequalities: $\mathcal{E}_3 > \mathcal{E}_{1,2,4}$ and $\mathcal{E}_1 \leqslant \mathcal{E}_4$, hold  independent of the value of $S$. Therefore, only three different cases are possible: (a) $\mathcal{E}_2<\mathcal{E}_1<\mathcal{E}_4<\mathcal{E}_3$, (b) $\mathcal{E}_1<\mathcal{E}_2<\mathcal{E}_4<\mathcal{E}_3$, and (c) $\mathcal{E}_1<\mathcal{E}_4<\mathcal{E}_2<\mathcal{E}_3$. Which case is realized depends on the value of the total spin $S$ in the ground state with $S_z=S$.
The energy $\mathcal{E}_1$ ($\mathcal{E}_2$) is the energy needed for an electron with spin up (down) which tunnels to the lowest available single-particle level (see Fig. \ref{Fig:XXX}). The energies $\mathcal{E}_4$
 and $\mathcal{E}_3$ are required for the tunneling of a spin-down electron to the lowest single-particle level available for an electron with spin up. The energy  $\mathcal{E}_3$ ($\mathcal{E}_4$) corresponds to the final excited state with the total spin $S-1/2$ ($S+1/2$). 

\begin{figure}[t]
	\centerline{\includegraphics[width=8cm]{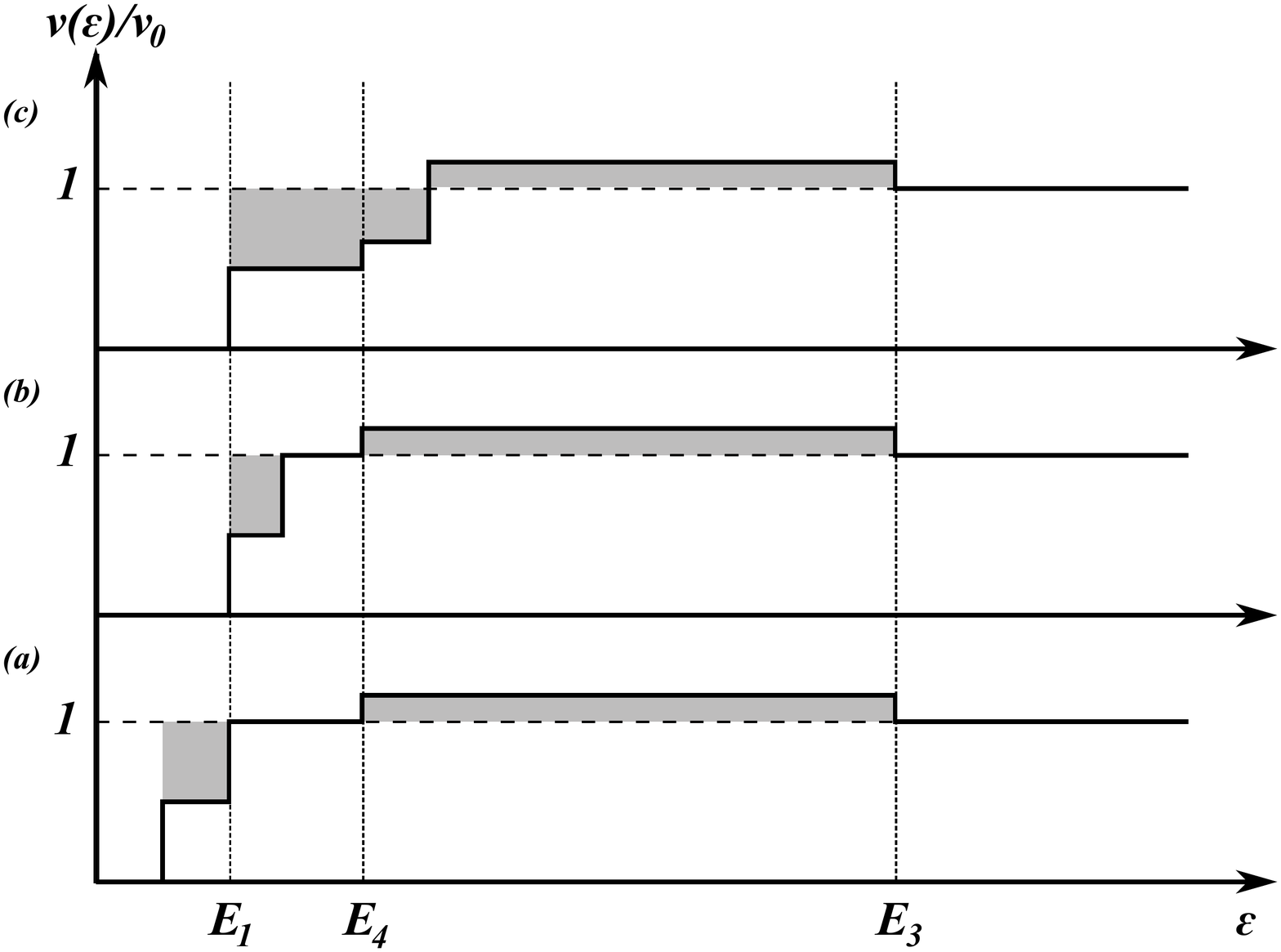}}
	\caption{The sketch of dependence of the tunneling density of states on energy at zero temperature. The shaded areas are equal (see text).}
	\label{Fig:Nu}
\end{figure}

As follows from Eq. \eqref{TDoS-3} the tunneling is possible only if an electron energy $\varepsilon$ exceeds $\min\{\mathcal{E}_1,\mathcal{E}_2,\mathcal{E}_3,\mathcal{E}_4\}$. Due to a finite value of the total spin in the ground state the tunneling of an electron is sensitive to its spin projection. Initially, only electrons with one spin projection can tunnel into the dot. For electrons with very large energies there is no dependence of  tunneling on their spin projection. The characteristic energy that separates such large energies is given by $\mathcal{E}_3$. 
The sum rule \eqref{eq:sumrule} restricts possible behavior of the tunneling density of states. For the Coulomb valley and at low temperatures the sum rule \eqref{eq:sumrule} forces the integral $\int d\varepsilon\, \nu(\varepsilon)$ to be independent of the exchange interactions. As we shall see below this leads to the existence of the maximum in the tunneling density of states. In the cases (a)  and (b) we find from Eq. \eqref{TDoS-3}:
\begin{equation}
	 \int\limits_{\varepsilon_1}^{\varepsilon_2} \frac{d\varepsilon}{\varepsilon_2-\varepsilon_1} \, \frac{\nu(\varepsilon)}{\nu_0} = \begin{cases}
		0, &  \varepsilon_{1,2} <\underline{\mathcal{E}_{12}} ,\\
		1/2, &  \underline{\mathcal{E}_{12}} \leqslant\varepsilon_{1,2} < \overline{\mathcal{E}_{12}} ,\\
		1, &  \overline{\mathcal{E}_{12}} \leqslant \varepsilon_{1,2} < \mathcal{E}_4 ,\\
		\frac{4S+3}{4S+2}, & \, \mathcal{E}_4 \leqslant \varepsilon_{1,2} < \mathcal{E}_3 ,\\
		1, & \mathcal{E}_3 \leqslant \varepsilon_{1,2} ,
	\end{cases}
	\label{eq:a&b}
\end{equation}
where $\underline{\mathcal{E}_{12}} = \min\{\mathcal{E}_1, \mathcal{E}_2\}$ and $\overline{\mathcal{E}_{12}} = \max\{\mathcal{E}_1, \mathcal{E}_2\}$. For the case (c) Eq. \eqref{TDoS-3} yields
\begin{equation}
	\int\limits_{\varepsilon_1}^{\varepsilon_2} \frac{d\varepsilon}{\varepsilon_2-\varepsilon_1} \, \frac{\nu(\varepsilon)}{\nu_0}  =  \begin{cases}
		0, & \, \varepsilon_{1,2} <\mathcal{E}_1 ,\\
		1/2, & \, \mathcal{E}_1 \leqslant \varepsilon_{1,2} < \mathcal{E}_4 ,\\
		\frac{S+1}{2S+1}, & \, \mathcal{E}_4 \leqslant\varepsilon_{1,2} < \mathcal{E}_2 ,\\
		\frac{4S+3}{4S+2}, & \, \mathcal{E}_2 \leqslant \varepsilon_{1,2} < \mathcal{E}_3 ,\\
		1, & \, \mathcal{E}_3 \leqslant \varepsilon_{1,2} . 
	\end{cases}
	\label{eq:c}
\end{equation}
Here $\nu_0 = 2\sum_\alpha \delta(\varepsilon-\epsilon_\alpha)$ denotes the density of states in the absence of interactions. Energies $\varepsilon_{1,2}$ are measured with  respect to $E_c-\mu$. The sketch of dependence of the tunneling density of states on energy at $T\ll \delta$ is shown in Fig. \ref{Fig:Nu}.

As follows from Eqs. \eqref{eq:a&b} and \eqref{eq:c} there is a step of height 1/2 between $\mathcal{E}_1$ and $\mathcal{E}_2$ in the envelope of the tunneling density of states for all three cases. However, for the ground state with the total spin $S \approx (J_\perp+J_z-\delta)/[2(\delta-J_z)]$ one can demonstrate that  $|\mathcal{E}_1 - \mathcal{E}_2| < J_z$, thus this 1/2 step will be smeared by temperature $T \gtrsim \delta$. The maximum in the tunneling density of states has the width of the order of $\mathcal{E}_3-\mathcal{E}_4 = J_\perp (2S+1) \sim J_\perp^2/(\delta-J_z)$. Therefore, for temperatures in the range $\delta \ll T \ll J_\perp^2/(\delta-J_z)$ one can expect that this maximum (of the relative height of $1/(2S) \sim [\delta-J_z]/J_\perp \ll 1$) survives. We note that such temperature regime exists for $\sqrt{\delta(\delta-J_z)} \ll J_\perp$ only. 

We emphasize that the zero temperature analysis demonstrates clearly that the tunneling density of states has only single maximum. There is no other extrema in contrast to findings of Ref. [\onlinecite{KiselevGefen}] based on perturbation theory in $J_\perp/J_z$.

\section{The tunneling density of states at $T\gg \delta$\label{Sec:TDOS_T}}

Now we analyze the behavior of the tunneling density of states at $T\gg \delta$. We neglect the effect of single-particle level fluctuations. We restrict our analysis below to the case $J_z\geqslant J_\perp$. We start from the integral representation \eqref{eq:K:fin} for the single-particle Green's function. In order to perform integration over $\phi_0$ and $h$ it is convenient to express it through the noninteracting single-particle Green's function. Then,
provided the condition $\mu \gg T\gg \delta$ holds we can approximate the noninteracting single-particle Green's function as
\begin{equation}
G_{0}(\tau) = -\sum_\alpha \frac{e^{(\epsilon_\alpha-\mu)\tau}}{1+e^{\beta (\epsilon_\alpha-\mu)}}\approx
-\frac{\pi T}{\delta} \frac{1}{\sinh(\pi T\tau)} .
\end{equation} 
Then, after integration over $\phi_0$, $h$ we obtain
\begin{widetext}
\begin{align}
\frac{\nu(\varepsilon)}{\nu_0} &= \frac{1+e^{-\beta\varepsilon}}{Z_C}  \sum_n e^{-\beta E_c(n-N_0)^2+\beta \mu n} 
\Biggl \{ \mathcal{F}\bigl ( \beta \Omega^\varepsilon_n, \beta (\delta-J_z)/4 \bigr )
+ \frac{\sqrt{\delta}}{16 Z_S \sqrt{(J_z-J_\perp)}}  \int\limits_{-\infty}^\infty d b\, \frac{e^{-b^2/[4\beta(J_z-J_\perp)]} e^{b/4}}{\cosh(b/4)\sinh(b/2)}
\notag \\
& \times \sum_{s,p=\pm} s p \, \mathbb{F}\left (\beta \Omega^\varepsilon_n - \frac{b}{2}, \frac{2bs+\beta\delta+p \beta J_\perp}{2\beta(\delta-p J_\perp)},\sqrt{\frac{\beta(\delta-p J_\perp)^2}{4(\delta-J_\perp)}},\frac{\sqrt{\beta(J_z-J_\perp)}}{2} \right )  
\Biggr \}
\label{eq:TDOS:int:1}
\end{align}
\end{widetext}
where $ \Omega^\varepsilon_n = \varepsilon+\mu - E_c(2n-2N_0+1)$. Here we introduce the following functions:
\begin{align}
\mathcal{F}(x,y)  =\int\limits_{-\infty}^\infty \frac{dt}{2 \cosh(\pi t)} e^{x (it+1/2)} e^{-y (t^2+1/4)}  
\end{align}
and
\begin{align}
\mathbb{F}(x,y,z,u)  =  \int\limits_{-\infty}^\infty \frac{dt\, e^{x (i t+1/2)}}{2\cosh(\pi t)}  \erfi\bigl (z(y-i t)\bigr ) e^{u^2 (t^2+1/4)} .
\end{align}
The quantity 
\begin{equation}
Z_C=\sum_n e^{-\beta E_c(n-N_0)^2+\beta\mu n} .
\end{equation}
denotes the contribution to the grand canonical partition function in the absence of exchange interaction and single-particle levels. The contribution due to exchange interaction is given by [\onlinecite{SLB}]
\begin{equation}
Z_S  = \left (\frac{\delta}{\delta-J_z}\right )^{1/2}e^{\frac{\beta J_\perp^2}{4(\delta-J_\perp)}} 
 \mathcal{F}_1\Bigl(\frac{\delta}{(\delta-J_\perp)}, \sqrt{\beta J_*}\Bigr ) ,
\end{equation}
where  $J_* = (\delta-J_\perp)(J_z-J_\perp)/(\delta-J_z)$ and 
\begin{equation}
\mathcal{F}_1(x,y) = \int\limits_{-\infty}^\infty d t \, \frac{\sinh(x y t)}{\sinh(y t)}\, e^{-t^2} .
\end{equation}

Using the following asymptotes of the function $\mathcal{F}(x,y)$:
\begin{equation}
\mathcal{F}(x,y) = \frac{1}{2} + \frac{1}{2} \begin{cases}
\tanh \frac{x-y}{2}, & y \ll 1  , \\
\erf (\frac{x-y}{2\sqrt{y}}), & y\gg 1 ,
\end{cases}
\label{eq:as:F}
\end{equation}
and the following relations $\mathcal{F}(-x,y)=e^{-x} \mathcal{F}(x,y)$ and
\begin{equation}
\partial_y \mathbb{F}(x,y,z,u) = \frac{2z}{\sqrt\pi} e^{z^2(y+1/2)^2} \mathcal{F}\bigl (x-2yz^2,\sqrt{z^2-u^2}\bigr )
\label{eq:as:F1}
\end{equation}
one can check that in the case of isotropic exchange, $J_z=J_\perp$, the result \eqref{eq:TDOS:int:1} coincides with the result obtained in Ref. [\onlinecite{BGK2}].  As follows from Eqs. \eqref{eq:as:F} and  \eqref{eq:as:F1} the functions $\mathcal{F}(x,y)$ and $\mathbb{F}(x,y,z,u)$ deviates only slightly from the Fermi function $f_F(x) =1/(1+\exp(x))$ for $y,z,u\ll 1$. Although $u$ is always small, the parameters $y$ and $z$ can be large for $J_\perp$ close to $\delta$ at temperatures $(\delta+J_\perp)^2/[4(\delta-J_\perp) \gg T \gg \delta$. 

\begin{figure}[t]
	\centerline{\includegraphics[width=8cm]{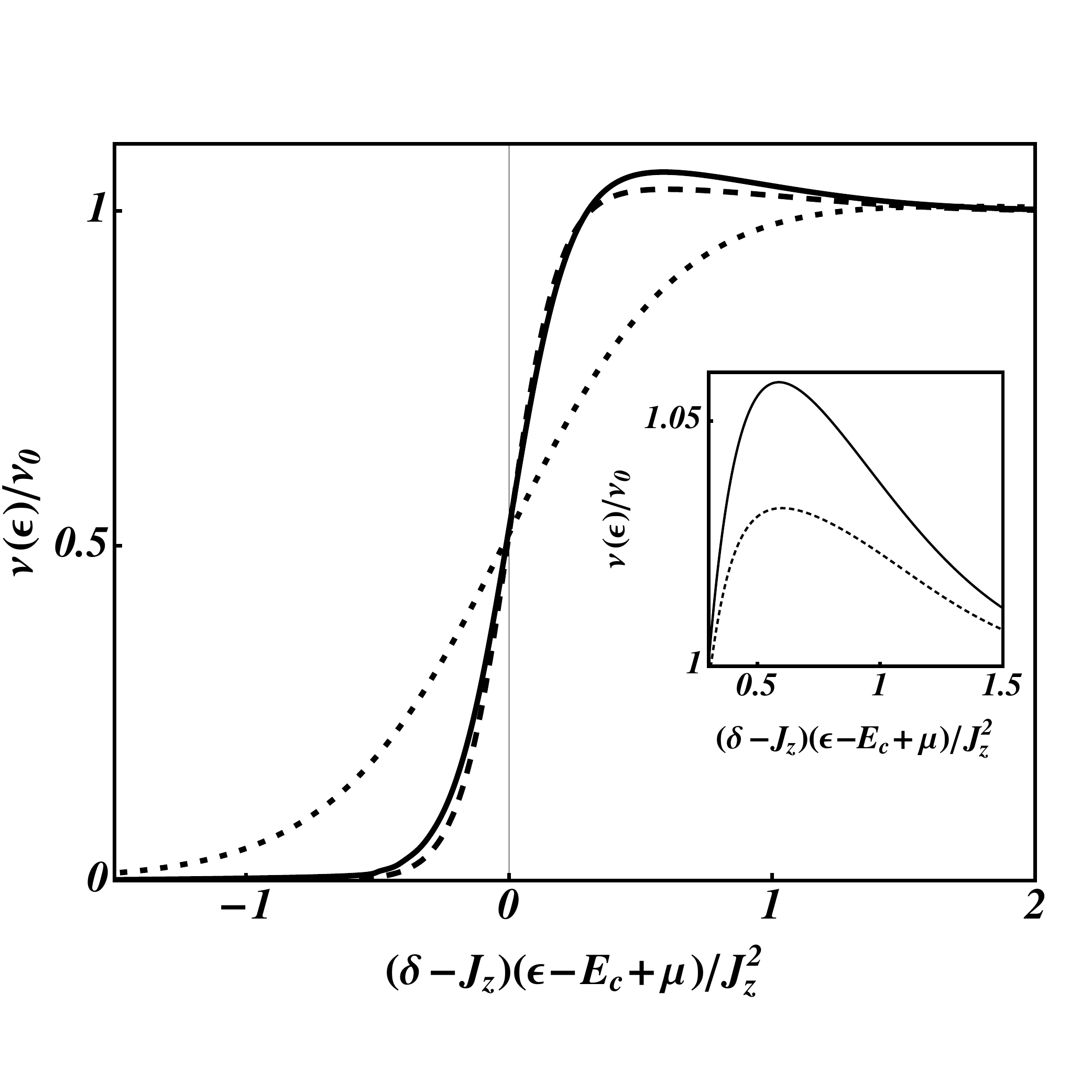}}
	\caption{The tunneling density of states in the Coulomb valley. The solid (dashed) curve corresponds to $J_z/\delta=0.92$, $J_\perp/\delta=0.85$ and $T/\delta=1.02$ ($T/\delta=3$). The dotted curve corresponds to 
$J_z/\delta=J_\perp/\delta=0.92$ and $T/\delta=1.02$.}
	\label{Fig:TDOS:Comp}
\end{figure}

The plot of the energy dependence of the tunneling density of states is shown in Fig. \ref{Fig:TDOS:Comp}. In accordance with zero temperature analysis there is the maximum of $\nu(\varepsilon)$ at non-zero values of $J_\perp$ and temperature $T \ll (\delta+J_\perp)^2/[4(\delta-J_\perp)$. We note that for the solid curve in Fig. \ref{Fig:TDOS:Comp} we choose precisely the same parameters as in Fig. 2 of Ref. [\onlinecite{KiselevGefen}]. There are no additional extrema contrary to conclusions of Ref. [\onlinecite{KiselevGefen}] based on perturbation theory in $J_\perp/J_z$.

\section{Conclusions\label{sec:dc}}

In this paper we studied the Hamiltonian which is an extension of the universal Hamiltonian to the case of uniaxial anisotropic exchange interaction.  Within this model we have derived exact analytic expression (see Eq. \eqref{TDoS}) for the tunneling density of states for arbitrary single-particle spectrum. For $(\delta-J_z) \ll  J_\perp<J_z$ we analyzed the energy dependence of the tunneling density of states for the equidistant single-particle levels at low ($T\ll \delta$) and high $(T\gg\delta$) temperatures. In both cases we demonstrated that in addition to non-monotonicities due to Coulomb blockade there is the maximum in the tunneling density of states at the characteristic energy of the order of $J_\perp^2/(\delta-J_z)$ (see Fig. \ref{Fig:TDOS:Comp}). The relative height of the maximum scales as $(\delta-J_z)/J_\perp$.  Qualitatively, in the case of anisotropic exchange the tunneling density of states has the same energy dependence as for the case of isotropic exchange. Our findings demonstrate that additional extrema in energy dependence of the tunneling density of states related with the anisotropic exchange interaction obtained in Ref. [\onlinecite{KiselevGefen}] on the basis of perturbative analysis in $J_\perp/J_z$ do not exist.

The most promising regime for experimental investigation of non-monotonicity due to exchange interaction in the tunneling density of states  is vicinity of the Stoner instability.  For example, one can perform the scanning tunneling microscopy in system of nanoparticles made of nearly ferromagnetic materials, e.g. Pd or Pt with Co or Ni impurities, various transition-metal alloys with dissolved Fe or Mn atoms, and rare-earth materials~[\onlinecite{Exp1}]. A very promising candidate could be nanoparticles made from YFe$_2$Zn$_{20}$ compound which has the exchange interaction $J \approx 0.94 \delta$  [\onlinecite{Exp2}]. 
For quantum dots fabricated in two-dimensional electron systems exchange interaction is typically not large, $J\lesssim \delta/2$ and, therefore, can be important only at $T\lesssim \delta$ [\onlinecite{QDLowT}]. However, recent experiments on two-dimensional strongly interacting electron systems in Si-MOSFET revealed the existence of electron droplets with finite spin of the order of $2$ at low temperatures and low densities [\onlinecite{RKTP}]. The well-known enhancement of electron-electron interaction in the triplet channel [\onlinecite{Finkelstein}] suggests that the physics of mesoscopic Stoner instability in disordered electron system [\onlinecite{Narozhny}] can be relevant for Si-MOSFET at low electron densities [\onlinecite{Future}].

\begin{acknowledgments}
We acknowledge useful discussions with Y. Gefen, I. Kolokolov, A. Kuntsevich, V. Pudalov, and A. Shnirman. The research was funded by Russian Science Foundation under the grant No. 14-02-00879.
\end{acknowledgments}

\end{document}